\documentclass{aastex701}

\usepackage{amsmath}

\begin{document}

\title{GRB 240205B: A Reverse Shock Detected in Rapid Response Radio Observations}

\author{S. I. Chastain}
\affiliation{Department of Physics and Astronomy, University of New Mexico, 210 Yale Blvd NE, Albuquerque, NM, 87106, USA}
\affiliation{Department of Physics \& Astronomy, Texas Tech University, PO Box 41051, Lubbock, TX, 79409, USA}
\email[show]{sarchast@ttu.edu}

\author{G. E. Anderson}
\affiliation{Australia Telescope National Facility, CSIRO, Space and Astronomy, PO Box 1130, Bentley, WA  6151, Australia}
\affiliation{Sydney Institute for Astronomy, School of Physics, The University of Sydney, NSW 2006, Australia}
\email{gemma.anderson@csiro.au}

\author{A. J. van der Horst}
\affiliation{Department of Physics, George Washington University, 725 21st St NW, Washington, DC, 20052, USA}
\email{ajvanderhorst@email.gwu.edu}

\author{L. Rhodes}
\affiliation{Trottier Space Institute at McGill, 3550 Rue University, Montreal, Quebec H3A 2A7, Canada}
\affiliation{Department of Physics, McGill University, 3600 Rue University, Montreal, Quebec H3A 2T8, Canada}
\email{lauren.rhodes@mcgill.ca}

\author{C. Morley}
\affiliation{International Centre for Radio Astronomy Research, Curtin University, GPO Box U1987, Perth, WA 6845, Australia}
\email{claire.l.morley@student.curtin.edu.au}

\author{A. Gulati}
\affiliation{Sydney Institute for Astronomy, School of Physics, The University of Sydney, NSW 2006, Australia}
\affiliation{CSIRO Space and Astronomy, PO Box 76, Epping, NSW 1710, Australia}
\affiliation{ARC Centre of Excellence for Gravitational Wave Discovery (OzGrav), Hawthorn, VIC 3122, Australia}
\email{agul8829@uni.sydney.edu.au}

\author{J. K.\ Leung}
\affiliation{David A. Dunlap Department of Astronomy and Astrophysics, University of Toronto, 50 St. George Street, Toronto, ON M5S 3H4, Canada}
\affiliation{Dunlap Institute for Astronomy and Astrophysics, University of Toronto, 50 St. George Street, Toronto, ON M5S 3H4, Canada}
\affiliation{Racah Institute of Physics, The Hebrew University of Jerusalem, Jerusalem 91904, Israel}
\email{jamesk.leung@utoronto.ca}

\author{T.~D.~Russell}
\affiliation{INAF - IASF Palermo, via Ugo La Malfa, 153, I-90146 Palermo, Italy}
\email{thomas.russell@inaf.it}

\author{S. D.\ Ryder}
\affiliation{School of Mathematical and Physical Sciences, Macquarie University, NSW 2109, Australia}
\affiliation{Astrophysics and Space Technologies Research Centre, Macquarie University, Sydney, NSW 2109, Australia}
\email{stuart.ryder@mq.edu.au}
\begin{abstract}

 Here we present broadband radio modeling of GRB 240205B, using observations with the Australia Telescope Compact Array (ATCA) and the South African MeerKAT radio telescope. Our observations include an automatically triggered early-time ATCA observation that began approximately 13 minutes after the gamma-ray signal and continued for 12 hours, resulting in the earliest detected GRB radio afterglow to date at $\sim35$ minutes post-burst. Following this initial detection, we conducted an extensive radio follow-up campaign for more than 5 months. Although the observations beyond one day post-burst are well described by a standard forward shock model, the observation before one day post-bust reveals an additional synchrotron component, which can be explained as the reverse shock. This component would have been missed without the automated ATCA rapid-response trigger. We find that a combined reverse and forward shock model in a stellar wind medium best describes the radio afterglow. We constrain the spectral breaks due to synchrotron self-absorption and the minimum electron energy, and we use the light-curve peaks to constrain the microphysical parameters. We put GRB 240205B in the context of the growing sample of GRBs with radio detections in the first hours after the gamma-ray trigger. Using our rapid response observation, we estimate the highest model independent constraint on a GRB minimum bulk Lorentz factor of $\Gamma\sim100$ at $\sim35$ minutes post burst. We also discuss future prospects of detecting similar long GRBs at centimeter wavelengths, as well as potential improvements to future strategies for targeting their radio afterglows.

\end{abstract}

\keywords{Gamma-ray Bursts}

\section{Introduction}
\label{sec:introduction}

Gamma-ray bursts (GRBs) are among the most powerful explosions in the Universe, appearing when a powerful jet is launched by either the death of a massive star \citep{1993ApJ...405..273W} or the collision of two compact objects including at least one neutron star \citep{1989Natur.340..126E}. These bursts are separated into two classes by their observed duration and spectral properties at gamma-ray energies. Those bursts longer than two seconds are called long GRBs and those shorter than two seconds are short GRBs, with long GRBs also being spectrally softer than short GRBs \citep{1993ApJ...413L.101K}. Although these properties are detector dependent, as a population, long and short GRBs tend to be associated with two kinds of phenomena. Long GRBs are expected to occur after the death of a massive star, while short GRBs occur in a binary merger involving at least one neutron star. Although, in recent times, there have been GRBs that counter these expectations \citep{2021MNRAS.503.2966R,2022Natur.612..223R,2022Natur.612..228T,yang22nat,rossi22,2024Natur.626..737L,yang24nat,dimple25}.

Following the initial prompt emission observed in gamma rays is an afterglow observable from TeV gamma rays down to radio wavelengths. This afterglow is the result of the GRB jet sweeping up material from the surrounding environment and forming a shock front. The forward shock propagates away from the jet into the external medium and is responsible for the bulk of the emission from the GRB afterglow \citep{rees92,sari97,piran99}. Some GRBs also have an observable reverse shock, which forms from the shock that propagates back into the jet. Although the reverse shock has been detected as an optical flash \citep{2000ApJ...542..819K,2000ApJ...536..195C,2005MNRAS.363...93Z,2013ApJ...776..120Y}, such signatures are difficult to catch, as they fade below detectability within minutes post-burst.
However, the reverse shock evolves more slowly at radio wavelengths, peaking, in some cases $\sim1$\,day post-burst \citep[e.g.][]{2014MNRAS.440.2059A}. Given the recent improvements in radio telescope sensitivity and the introduction of rapid-response observing systems, more reverse shock detections have been made at early times in the radio band 
\citep[e.g.][]{kulkarni99,frail00grb991216,berger03grb020405,soderberg06,2014MNRAS.440.2059A,2024ApJ...975L..13A,2014MNRAS.444.3151V,perley14grb130427a,laskar13grb130427a,laskar16grb160509a,laskar19grb181201a,laskar19grb190114c,lamb19,2020MNRAS.496.3326R,fong21,2023NatAs...7..986B}.

Detections of the reverse shock at radio wavelengths is still uncommon due to their rapid evolution, with few examples of high temporal and spectral sampling of this component \citep[e.g.][]{2023NatAs...7..986B}.
However, detecting the reverse shock emission is crucial for providing insight into jet properties such as its particle content and magnetization \citep[e.g.][]{laskar19grb190114c}, the structure of the surrounding environment \citep[e.g.][]{2014MNRAS.444.3151V}, and the initial Lorentz factors of the outflow \citep[e.g.][]{2024ApJ...975L..13A}.
In order to build the sample of radio-detected reverse shocks, we conceived the PanRadio GRB program \citep{2025arXiv251107644L}, which utilises the Australia Telescope Compact Array (ATCA) to perform automatically triggered (rapid-response) observations \citep{anderson21} and long-term multi-frequency radio monitoring of all long GRBs in the Southern Hemisphere.
Using the ATCA rapid-response mode, our team has obtained the earliest radio detections of GRBs, including the detection of the reverse shock from short GRB 230217A \citep[][triggered observation beginning at 32 minutes]{2024ApJ...975L..13A} and the detection of a radio flare from short GRB 231117A due to a violent shell collision \citep[][triggered observation beginning at 1.3 hours]{2025arXiv250814650A}. Even non-detections obtained from radio rapid-response observations can provide vital insight into GRB jet microphysics \citep{anderson21,chastain24}.


In this work, we present the results of our observations of GRB 240205B. 
The ATCA rapid-response observation began just 13 minutes after the initial burst. We detected the GRB afterglow in this initial observation, resulting in the earliest radio detection of a GRB afterglow to this date in a span of 13 to 56 minutes post-burst. As part of the PanRadioGRB program, we continued taking observations of this burst at multiple frequencies out to 161 days post-burst. While our radio light curve beyond one day post-burst is well described by a standard forward shock model, the rapid-response observation reveals an additional synchrotron component at early times (less than 12 hours post-burst). We therefore fit a combined reverse plus forward shock model to all of the observations. In Section~\ref{sec:observations}, we describe the observing strategy and observations.
In Section~\ref{sec:modeling}, we describe the modeling of the radio afterglow.
In Section~\ref{sec:discussion}, we discuss the implications of our model for future low frequency observations. Additionally, we put our constraints on the physical parameters and on the minimum bulk Lorentz factor, $\Gamma_{min}$, in context with other GRBs. Finally in Section~\ref{sec:conclusions}, we provide a summary and briefly review the implications of our findings on future studies.

\section{Observations}
\label{sec:observations}
An overview of all the radio observations of GRB 240205B, including the observing frequencies, are shown in Table~\ref{tab:obs}. After the initial observation, nine additional observations were carried out in a quasi-logarithimic fashion out to 161 days post-burst. Up to around 23 days after the burst, observations were carried out using the 4 cm and 15 mm dual receivers which had central frequencies of 5.5/9\,GHz and 16.7/21.2\,GHz, respectively, each with a 2 \,GHz bandwidth. After 23 days post-burst, observations at either 1.3 GHz (MeerKAT) or 1.6 GHz (ATCA), and 3.1 GHz (MeerKAT) or 2.6 GHz (ATCA), were taken, and observations at 16.7 GHz and 21.2 GHz were discontinued due to a combination of consistently poor observing conditions and an expected fading of the afterglow from higher frequencies. 
We also obtained a single MeerKAT epoch using the Ultra High Frequency (UHF) centred at 0.8 GHz 
around 23 days post-burst. 

\subsection{ATCA}
ATCA is a six-dish interferometer with a maximum baseline length of 6\,km based in New South Wales, Australia. 
Our observations of GRB 240205B were taken with a variety of array configurations, which are frequently changed 
so that ATCA can offer variation in \textit{uv}-coverage over the course of a year. 
Our observations were taken as part of the PanRadio GRB program (C3542, PI: Anderson), which uses the rapid-response mode \citep{anderson21} to automatically trigger observations of a GRB 
after it is detected by the Burst Alert Telescope \citep{2005SSRv..120..143B} on the {\it Neil Gehrels Swift Observatory} \citep[hereafter {\it Swift-BAT};][]{2004ApJ...611.1005G}. Further observations are then manually scheduled at a variety of radio frequencies to monitor the long term evolution of any associated radio afterglow. 

On February 5, 2024, at 22:13:06 UTC, \textit{Swift-BAT} detected a long GRB \citep{2024GCN.35683....1M}. Upon receiving the alert, the rapid-response observations
began 13 minutes later using the 4\,cm receiver (central frequencies of 5.5 and 9 GHz), observing for a total of 12 hours. Observations began with 9.8 minutes of set-up scans using the complex gain calibrator, PKS B2333-528. Observations then cycled between observing the complex gain calibrator for two minutes, followed by GRB 240205B for 20 minutes. A six-minute observation of the standard flux and bandpass calibrator PKS B1934-638 was taken when it became visible above the horizon at approximately seven hours after observations began. 

During the first two observations, the ATCA correlator was configured for spectral line observations (referred to as the Zoom mode). As a result, each 2,048 MHz band was configured with 33 channels with a width of 64 MHz and 2,049 channels of width 0.031 MHz centered at 6,236 MHz and 9,736 MHz, instead of the standard configuration of 2,048 channels of width 1 MHz. For these observations, the zoom channels were discarded, and only the two bands of 33 channels were used. Despite the lowered spectral resolution, the observations were still of sufficient quality to obtain a detection. All other ATCA observations were taken using the standard correlator configuration. The correlator mode and array configuration for each ATCA observation is included in Tables~\ref{tab:obs} and \ref{tab:uvmeas}. The reported errors in the tables are statistical only.

\begin{figure*}
	\centering
	\includegraphics[width=\textwidth]{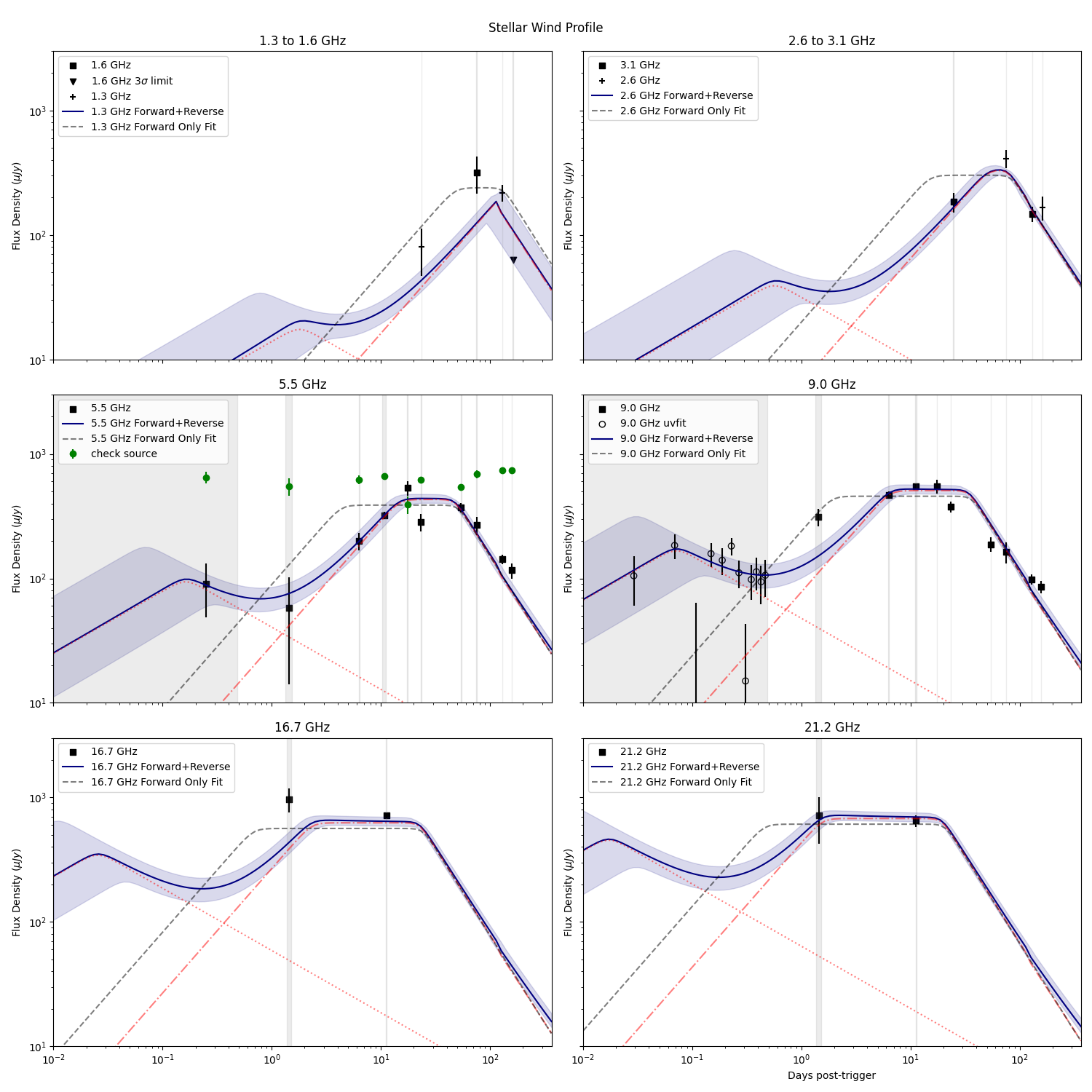}
	\caption{GRB 240205B afterglow light curves for each radio observing frequency. The open circles show data from the \textit{uv}-fitting of the initial 9~GHz observation. The solid lines show the best fit for a combined forward plus reverse shock model for a stellar wind density profile, with a reduced $\chi^2$ of 3.0 with 36 degrees of freedom (DOF); the dashed lines show a fit using the forward shock only, which has a reduced $\chi^2$ of 6.2 with 38 DOF. 68\% confidence intervals for the combined fits are indicated by the shaded regions around the lines. The dotted lines indicate the reverse shock component of the combined fit and the dash-dotted lines indicate the forward shock component of the combined fit. The vertical shaded regions indicate the duration of the observation. The statistical error of each flux measurement is added in quadrature with the RMS noise to give the errors shown by the error bars.}
	\label{fig:lc}
\end{figure*}

\begin{figure*}
	\centering
	\includegraphics[width=\textwidth]{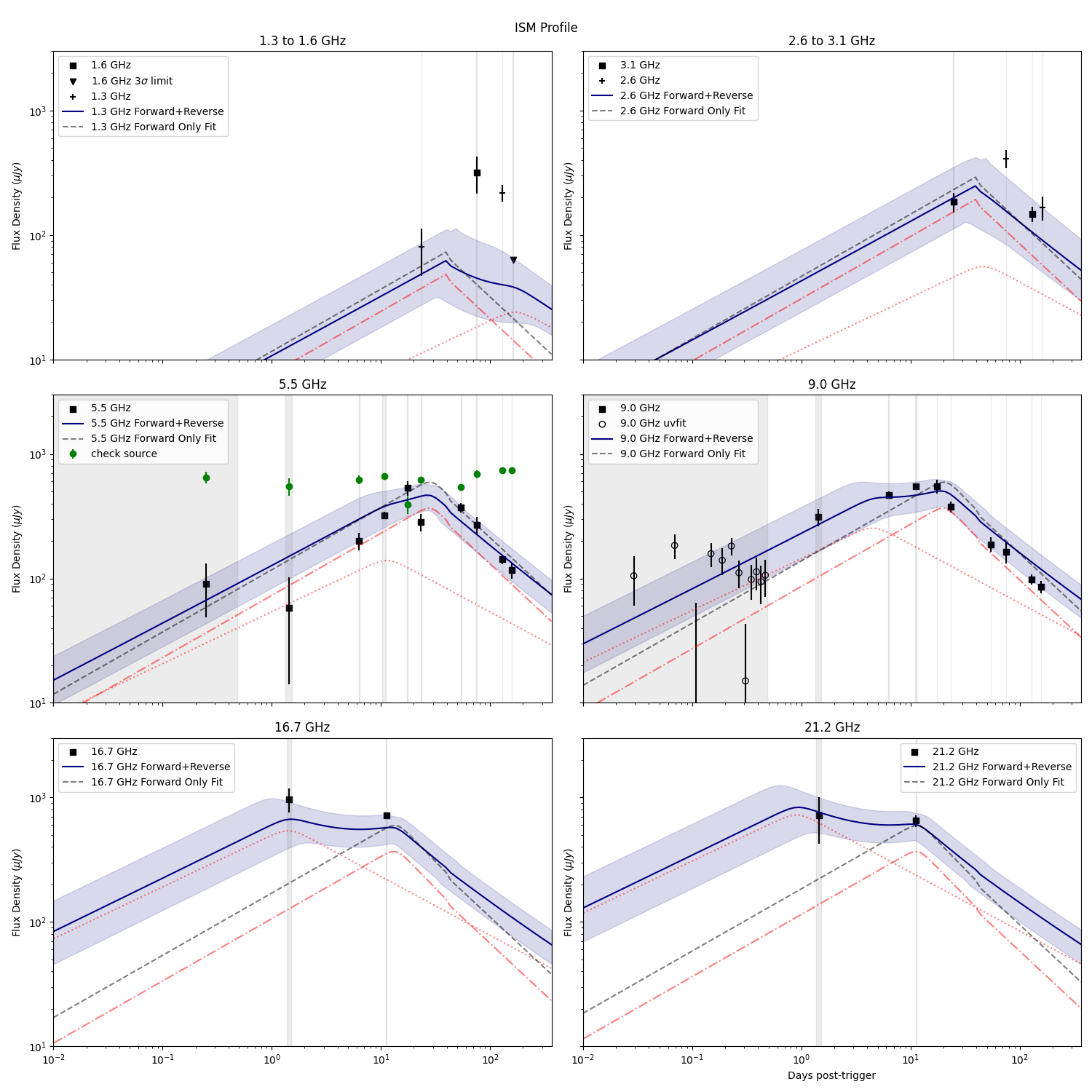}
	\caption{Same as for Figure~\ref{fig:lc} except assuming a uniform density profile with the combined forward plus reverse shock model resulting in a reduced $\chi^2$ of 5.4 with 36 DOF; the dashed lines show a fit using the forward shock only, which has a reduced $\chi^2$ of 7.6 with 38 DOF. 
    }
	\label{fig:lcISM}
\end{figure*}

The ATCA data were imported from the RPFITS format into the Common Astronomy Software Applications \citep[\textsc{CASA};][]{2022PASP..134k4501C} software using the \textit{importatca} task. The data was then flagged for radio-frequency interference (RFI) and calibrated using standard \textsc{CASA} tasks. The calibrated image was imaged using \textit{tclean} with the standard gridder and deconvolution algorithm. The source fitting of the detected GRB afterglow was performed in the CARTA (Cube Analysis and Rendering Tool for Astronomy) software (DOI 10.5281/zenodo.3377984 –  https://cartavis.github.io). The background flux, the source position, the ellipticity, and source flux were all free parameters. The resulting residuals and models for the resulting fits were visually examined to ensure the source was properly fit. The statistical errors from the fit, reported in Tables \ref{tab:obs}, were added in quadrature to the RMS noise of the image. These calculated errors were the errors used in the model fitting. At 5.5 GHz, we included a check source to ensure that there were no systematic issues with the flux calibration. This source was within the same field of view as the GRB and appeared to be a point source in all of the observations. This check source is shown in the 5.5 GHz light curves in Figure~\ref{fig:lc} and Figure~\ref{fig:lcISM}. Comparing the check source to the afterglow shows no significant correlated variability providing confidence that our measured afterglow fluxes are not varying due to systematic errors.

In order to measure the flux evolution at 9 GHz within the first observation, we performed \textit{uv}-fitting. We performed three rounds of \textit{uv}-subtraction on all of the sources in the image except for the afterglow, in order to more accurately model and remove the other sources in the field. We then imaged the field to confirm that the GRB afterglow was the only source remaining. Using \textsc{CASA}'s \textit{uvmodelfit} task, we fit a point source model to the location of the GRB afterglow allowing the position to vary. 
The first measurement has a central time of 35 minutes post-burst, which is the earliest published radio detection of a GRB to-date \citep[with the previous record holder being the short GRB 230217A at 1 hour;][]{2024ApJ...975L..13A}. 
These measurements are shown in Table~\ref{tab:uvmeas} and as open circles in Figures~\ref{fig:lc} and \ref{fig:lcISM}.

\begin{table}
	\caption{Radio observations of GRB 240205B with ATCA and MeerKAT. Reported 1-$\sigma$ uncertainties are statistical only.}
	\label{tab:obs}
	\begin{tabular}{llllll}
		\hline
		Days      & Freq. & $F_{pk}$  & RMS noise & Array & Telescope\\
		post-burst & (GHz) & ($\mu Jy$) & ($\mu Jy/{\rm{beam}}$) & Configuration$^a$ &\\
		\hline
		0.009-0.487 & 5.5 & $90\pm27$ & 30 & EW367 &  ATCA$^b$ \\
		0.009-0.487 & 9.0 & $103\pm24$ & 11 & EW367 &  ATCA$^b$ \\
		1.338-1.521 & 5.5 & $58\pm32$ & 43 & EW367 &  ATCA$^b$\\
		1.338-1.521 & 9 & $312\pm43$ & 29 & EW367 &  ATCA$^b$\\
		1.367-1.511 & 16.7 & $969\pm202$ & 61 & EW367 &  ATCA$^b$\\
		1.367-1.511 & 21.2 & $714\pm216$ & 192 & EW367 &  ATCA$^b$\\
		6.260-6.378 & 5.5 & $199\pm10$ & 30 & 6A & ATCA \\
		6.260-6.378 & 9 & $467\pm25$ & 21 & 6A & ATCA\\
		11.057-10.341 & 5.5 & $320\pm17$ & 12 & 6A & ATCA\\
		11.057-11.341 & 9 & $550\pm18$ & 11 & 6A & ATCA\\
		11.088-11.320 & 16.7 & $712\pm27$ & 20 & 6A & ATCA\\
		11.166-11.320 & 21.2 & $648\pm54$ & 45 & 6A & ATCA\\
		17.369-17.528 & 5.5 & $532\pm53$ & 48 & 6A & ATCA\\
		17.369-17.528 & 9 & $549\pm54$ & 42 & 6A & ATCA\\
		23.494-23.538 & 0.82 &  & 25 & &  MeerKAT\\
		23.639-23.682 & 1.3 & $80\pm26$ & 20 &  & MeerKAT\\
		24.644-24.687 & 3.1 & $185\pm26$ & 22 &  & MeerKAT\\
		23.198-23.489 & 5.5 & $284\pm33$ & 33 & 6A & ATCA\\
		23.198-23.489 & 9 & $376\pm5$ & 38 & 6A & ATCA \\
		54.073-54.427 & 5.5 & $370\pm23$ & 21 & 6A & ATCA\\
		54.073-54.427 & 9 & $188\pm19$ & 17 & 6A &ATCA \\
		74.906-75.344 & 1.6 & $320\pm85$ & 62 & 6A &ATCA \\
		74.906-75.344 & 2.6 & $413\pm53$ & 46 & 6A &ATCA \\
		74.906-75.344 & 5.5 & $268\pm32$ & 32 & 6A & ATCA\\
		74.906-75.344 & 9 & $164\pm20$ & 25 & 6A & ATCA\\
		130.024-130.059 & 1.3 & $219\pm23$ & 26 & & MeerKAT$^c$ \\
		130.024-130.059 & 3.1 & $148\pm15$ & 15 & & MeerKAT$^c$ \\
		128.744-129.192 & 5.5 & $143\pm10$ & 7 & 6D & ATCA\\
		128.744-129.192 & 9 & $98\pm7$ & 6 & 6D & ATCA\\
		161.543-161.988 & 1.6 & & 21 & 6D & ATCA\\
		161.543-161.988 & 2.6 & $167\pm34$ & 15 & 6D & ATCA\\
		157.593-157.593 & 5.5 & $116\pm14$ & 9 & 6D & ATCA\\
		157.593-157.593 & 9 & $86\pm7$ & 7 & 6D & ATCA\\
		\hline
        \multicolumn{6}{l}{\small $^a$\url{https://www.narrabri.atnf.csiro.au/operations/array_configurations/configurations.html}} \\
        \multicolumn{6}{l}{\small $^b$Zoom} \\
        \multicolumn{6}{l}{\small $^c$Sub-array}
	\end{tabular}
\end{table}
\begin{table}
	\caption{Results from \textit{uv}-fitting data from the first radio observations of GRB 240205B with ATCA. Reported 1-$\sigma$ uncertainties are statistical only.}
	\label{tab:uvmeas}
	\begin{tabular}{lllll}
		\hline
		Days      & Freq. & $F_{pk}$  &  Array & Telescope\\
		post-burst & (GHz) & ($\mu Jy$) & Configuration$^a$ &\\
		\hline
		0.009-0.039 & 9 & $105\pm45$ &   EW367 & ATCA$^b$\\
		0.039-0.071 & 9 & $184\pm41$ &   EW367 &  ATCA$^b$\\
		0.071-0.102 & 9 & $-7\pm71$ &   EW367 &  ATCA$^b$\\
		0.102-0.134 & 9 & $158\pm34$ &   EW367 &  ATCA$^b$\\
		0.134-0.165 & 9 & $140\pm34$ &  EW367 &  ATCA$^b$\\
		0.165-0.196 & 9 & $182\pm30$ &  EW367 &  ATCA$^b$\\
		0.197-0.228 & 9 & $111\pm28$ &  EW367 &  ATCA$^b$\\
		0.228-0.259 & 9 & $15\pm28$ &  EW367 &  ATCA$^b$\\
		0.260-0.291 & 9 & $98\pm31$ &  EW367 &  ATCA$^b$\\
		0.298-0.330 & 9 & $113\pm33$ &  EW367 &  ATCA$^b$\\
		0.330-0.361 & 9 & $18\pm28$ &  EW367 &  ATCA$^b$\\
		0.361-0.393 & 9 & $92\pm29$ &  EW367 &  ATCA$^b$\\
		0.393-0.424 & 9 & $109\pm30$ &  EW367 &  ATCA$^b$\\
		0.424-0.456 & 9 & $94\pm32$ & EW367 &  ATCA$^b$\\
		0.456-0.487 & 9 & $106\pm35$ &  EW367 &  ATCA$^b$\\
		\hline
        \multicolumn{5}{l}{\small $^a$\url{https://www.narrabri.atnf.csiro.au/operations/array_configurations/configurations.html}} \\
        \multicolumn{5}{l}{\small $^b$Zoom} 
	\end{tabular}
\end{table}
%
\subsection{MeerKAT}

MeerKAT is a 64-dish radio interferometer based in the Karoo Desert in South Africa. We were awarded Director's Discretionary Time on MeerKAT to observe the afterglow of GRB 240205B at central frequencies of 0.82, 1.28, and 3.06\,GHz  (DDT-20240228-LR-01, PI: Rhodes). The awarded time included an observation in all three bands using the full array starting at 23.5\,days after the gamma-ray trigger, and a second observation at 130\,days at 1.28 and 3.06\,GHz using two sub-arrays. For both observations, J1939-6342 was used as the bandpass and flux calibrator, and J2329-4730 was used as the complex gain calibrator. For all observations, the on-source time was 45 minutes at all frequencies. The data were reduced using \textsc{oxkat} \citep{2020ascl.soft09003H}, a series of python scripts that perform flagging, calibration (both first generation and phase-only self-calibration) and imaging of MeerKAT measurement sets. Both 1.28 and 3.06\,GHz observations resulted in detections, but no counterpart was detected at 0.82\,GHz in the first epoch. 
All results are given in Table \ref{tab:obs}. The reported errors in Table~\ref{tab:obs} are statistical only. The source measurements and errors used in the model fitting were calculated in the same way as in the ATCA data, by using CARTA and adding the statistical error in quadrature with the RMS noise. The observations marked ``sub-array'' in Table~\ref{tab:obs} were taken with MeerKAT configured into a sub-array with half of the elements observing at 1.3\,GHz and half of the elements observing at 3.06\,GHz.

\section{Modeling}
\label{sec:modeling}

GRB afterglows are typically modeled as synchrotron emission from electrons accelerated by strong shocks \citep{1998ApJ...497L..17S,1999ApJ...523..177W}. Synchrotron self-Compton effects can play a role at X-ray and higher energies; however, at radio frequencies, synchrotron self-absorption needs to be taken into account. The evolution of the synchrotron spectra is governed by the evolution of the shocks, for which one typically uses relativistic blast wave solutions \citep{1976PhFl...19.1130B} at early times and non-relativistic blast wave solutions \citep{1946JApMM..10..241S,1950RSPSA.201..159T} at later times \citep[months to years;][]{2000ApJ...537..191F}. 
These analytical blast wave solutions can be used to produce smoothly broken power-law models for the observed flux density as a function of frequency and time of the form $F \propto t^{\alpha} \nu^{\beta}$ for the forward shock \citep{2002ApJ...568..820G,2007PhDT........72V} and the reverse shock propagating back into the jet \citep{2000ApJ...542..819K,2000ApJ...536..195C,2013ApJ...776..120Y,2005MNRAS.363...93Z}.


We fit all the radio data of GRB 240205B from our observing campaign simultaneously. Observations of GRB 240205B in the optical and X rays were made difficult due to Sun constraints, resulting in very few measurements. Therefore, we only model the radio observations presented here. In examining our radio light curves, it is not possible to describe the behavior of the radio light curves with only one emission component. This is particularly true of the 9~GHz light curve, which spans four orders of magnitude in time. The 9 GHz light curve, shown in Figure~\ref{fig:lc}, starts with a rise and decay, followed by another rise and decay. Within the standard forward shock framework, there is no perceivable situation in the radio regime in which this light curve behavior can be explained by one forward shock. Therefore, we adopt a model that includes two emission components: a forward shock and a reverse shock. A comparison between a forward shock only model and a combined forward + reverse shock model is shown in both Figures~\ref{fig:lc} and~\ref{fig:lcISM}. Whenever a strong forward shock is formed, a reverse shock should also be formed, but observing significant emission from the latter depends on the physical conditions at the front of the jet \citep[e.g.,][]{2000ApJ...542..819K}.

We use a forward shock model derived from \cite{2007PhDT........72V} and a reverse shock model as described in \cite{2014MNRAS.444.3151V}. Since we are using radio data, we can in principle constrain the characteristic synchrotron frequency associated with the minimum energy of the accelerated electrons ($\nu_m$), the synchrotron self-absorption frequency ($\nu_a$), and the peak flux of the synchrotron spectrum ($F_0$). We model the spectral energy distribution as a double smoothly broken power law with the flux normalization and the spectral breaks evolving as single power laws over time. For these models, we use the expected spectral and temporal slopes given in \cite{2007PhDT........72V} and \cite{2014MNRAS.444.3151V}. In our modeling, we account for the transition when $\nu_m$ meets $\nu_a$, but, in our model, this transition is not smoothly broken, as can be seen at around 100 days in the 1.3 to 1.6 GHz light curve in Figure~\ref{fig:lc}. The available data do not allow us to constrain the power-law index $p$ of the electron energy distribution, so we set $p=2.2$ \citep[e.g.,][]{Kirk2000,Achterberg2001}. In our modeling, we consider two different scenarios for the density $\rho$ profile of the ambient medium as a function of radius $R$ ($\rho \propto R^{-k}$), i.e., a uniform density profile ($k=0)$ and a stellar wind structure ($k=2$), as is more common around long GRBs. The forward shock model has the following independent variables, parameters, and given values:
\begin{equation}
F_{forward}(t,\nu;F_{0,fwd},\nu_{a,0,fwd},\nu_{m,0,fwd}|p=2.2,k\in(0,2)) 
\end{equation}
For the forward shock only modeling, we simply fit this equation, which is a double smoothly broken power law evolving over time according to the standard models. For the reverse shock we use:
\begin{equation}
F_{reverse}(t,\nu;F_{0,rev},\nu_{a,0,rev}|p=2.2,k\in(0,2),\nu_{m,0,rev}=10^{12}) 
\end{equation}
Since we cannot constrain the second break in the reverse shock, we set $\nu_{m,0,rev}$, the initial value for $\nu_m$ in the reverse shock, to a very high value in order to remove it from consideration. For our combined forward and reverse shock model, we simply add these two components:
\begin{equation}
\begin{split}
F_{total} = \\
F_{forward}(t,\nu;F_{0,fwd},\nu_{a,0,fwd},\nu_{m,0,fwd}|p=2.2,k\in(0,2)) \\
+ F_{reverse}(t,\nu;F_{0,rev},\nu_{a,0,rev}|p=2.2,k\in(0,2),\nu_{m,0,rev}=10^{12})
\end{split}
\end{equation}

This analytic approach to modeling the observed radio data gives us a handle on the constraints we can set on the spectral parameters. Since we do not have multi-wavelength data available, an analysis that gives us all the macro- and micro-physical parameters of the blast wave and its environment is going to be limited. 

The fitting is done in Python using curve\_fit from SciPy \citep{2020SciPy-NMeth}. The code used to fit the data, and the data itself, are available at \url{https://github.com/dentalfloss1/grb240205b}.
We will first discuss our considerations for modeling the reverse shock, followed by our full modeling results for both the forward and reverse shock.


\subsection{Modeling the Reverse Shock}

By time-splitting and \textit{uv}-fitting the first 12-hour long observation at 9~GHz, we build a light curve with excellent temporal coverage in the radio at these early times. We are unable to do the same with the 5.5 GHz observations due to a nearby bright, partially resolved source in the field that cannot be fully removed from the visibility data. The middle-right panels of Figures~\ref{fig:lc} and~\ref{fig:lcISM} show the 9~GHz light curve with each data point in the initial 12-hour observation representing $\sim$45 minutes. This first 12 hours shows a potential rise and decaying component $<1$\,day post-burst, although not statistically significant. This early-time emission is poorly modeled by a forward-shock only model, and warrants the inclusion of a reverse shock, albeit with some fixed parameters due to low signal-to-noise. 

While the uncertainties on the \textit{uv}-fitted data in the first 12 hours are relatively large, there appears to be a declining trend in the observed flux that follows a potential rise in the first one or two hours. This rise is not statistically significant, as can be seen from the large uncertainties on the early light curve slope. The decay is consistent with the expected temporal slope of $\sim-0.5$ for a reverse shock with either a homogenous or stellar wind type profile with the observing frequency being above $\nu_a$ and below $\nu_m$. In this scenario, the potential early turn-over from rise to decay would be consistent with $\nu_a$ passing through the observing band. Since our data is not able to constrain multiple breaks in the reverse shock, we fixed the spectral breaks due to $\nu_c$ and $\nu_m$ to very high values in order to only fit the spectral break due to $\nu_a$, which is the peak for the reverse shock. 

It should be noted that in this ordering of the characteristic frequencies, the expected temporal slope is almost the same for a thick (relativistic) or a thin (non-relativistic) shell model, so our data set cannot distinguish between the two models. With the lack of early-time optical light curves from instruments such as the \textit{Swift}-Ultra-violet Optical Telescope \citep{2005SSRv..120...95R} or ground-based robotic optical telescopes, we are unable to further constrain the physics of the reverse shock. Nevertheless, as can be seen in Figure~\ref{fig:lc}, a reverse shock component is necessary to account for the early-time detections of the afterglow at 9 GHz and to a lesser extent at 5.5 GHz.

\subsection{Full Modeling Results}

After establishing the model to describe the very early-time data, we fit the entire data set simultaneously with a forward plus reverse shock model. We fit the data for this model propagating into a stellar wind medium and a homogeneous medium, and the results are shown in Figures~\ref{fig:lc} and \ref{fig:lcISM}, respectively. Table~\ref{tab:redchisq} shows the the reduced $\chi^2$ values of the fitted models, which are also compared to the poorer forward shock only model for completeness. There are two clear differences in the model fits. The first noticeable difference is that the ISM model fit shows an extremely bright and late reverse shock, which fails to provide a good fit to the 9 GHz \textit{uv}-fitted data from the first twelve hours. Additionally, the 1.3 to 1.6 GHz data are significantly under-predicted in this model, which highlights the value of multi-frequency observations in the radio regime. The wind medium with a forward and reverse shock is clearly favored. There are some measurements that do not fit this model, observation two at around 1.5 days post-burst, in particular. However, this can be explained, at least in part, by significant scintillation in the light curves, potentially out to tens of days, which has been observed in several GRB afterglows \citep[e.g.,][]{2000ApJ...537..191F,2003ApJ...590..992F}. Based upon the better reduced chi-squared values and the visibly better fit when considering the multi-frequency observations, we conclude that a model with a stellar wind medium leads to a better fit to the data.
\begin{table}
		\caption{Reduced $\chi^2$ values for the fitted models.}
		\centering
	\label{tab:redchisq}
	\begin{tabular}{|l|ll|}
        \hline
		Model     &  Reduced $\chi^2$ & DOF   \\
        \hline
        Forward Shock Only, Stellar Wind & 6.2 & 38 \\
        Forward Shock Only, ISM & 7.6 & 38 \\
        Forward + Reverse Shock, Stellar Wind & 3.0 & 36 \\
        Forward + Reverse Shock, ISM & 5.4 & 36 \\
        \hline
    \end{tabular}
\end{table}
For the forward shock, we are able to constrain the peaks and slope changes of the forward shock component in the light curves due to $\nu_a$ and $\nu_m$ passing through the observing band. For the fits with a homogeneous ambient medium, the self-absorption break $\nu_a$ does not evolve over time. When compared to the stellar wind case, this results in the spectral break due to $\nu_m$ reaching $\nu_a$ much earlier as it evolves to lower frequencies over time. The end result is that the low-frequency light curves for the homogeneous medium tend to peak earlier and have lower flux than what would be expected for a stellar wind environment. We can see these features when comparing the model of the $1.3-1.6$ GHz light curve in Figure~\ref{fig:lc} with Figure~\ref{fig:lcISM}. 

From our best-fitting model in a wind medium, we constrain the spectral breaks for the reverse and forward shocks. For the reverse shock at these wavelengths, we interpret the potential peak in the 9~GHz light curve at around six days as being due to $\nu_{a,0,fwd}$ passing through the observing frequency. Using a smoothly-broken power-law model, we find that at 0.05~days, $\nu_{a,0,rev}=11.\pm5.0$ GHz, and that the flux normalization, defined by $\nu_{m,0,rev}$ in this model, is $F_{0,rev}=0.28\pm0.15$~mJy. These parameters are not well constrained since the significance of this peak is low.

For the forward shock, we were able to constrain two of the spectral breaks, $\nu_{a,0,rev}$ and $\nu_{m,0,rev}$, as well as the flux normalization, $F_0$. We are not able to constrain the cooling break $\nu_c$ using our data, since it is typically situated in the optical to X-ray regime. We fixed the normalization time to be at 1 day, and find that at that time $\nu_{a,0,fwd}=27.7\pm1.9$ GHz, $\nu_{m,0,fwd}=(1.80\pm0.16)\cdot 10^3$ GHz, and $F_{0,fwd}=3.00\pm0.14$ mJy. The implications of these constraints will be discussed in the next section.


\section{Discussion}
\label{sec:discussion}
GRB 240205B offers a unique look at the impact of very early- to late-time radio observations over a broad range of radio frequencies. This dataset sheds light on the ambient medium density structure of this GRB, and the early detections provide a promising outlook for detecting and measuring the reverse shock in future GRBs. Here we discuss some further implications of the observations presented in this paper.

\subsection{Low Frequency Observations}

\begin{figure}
	\centering
    \includegraphics[width=0.48\textwidth]{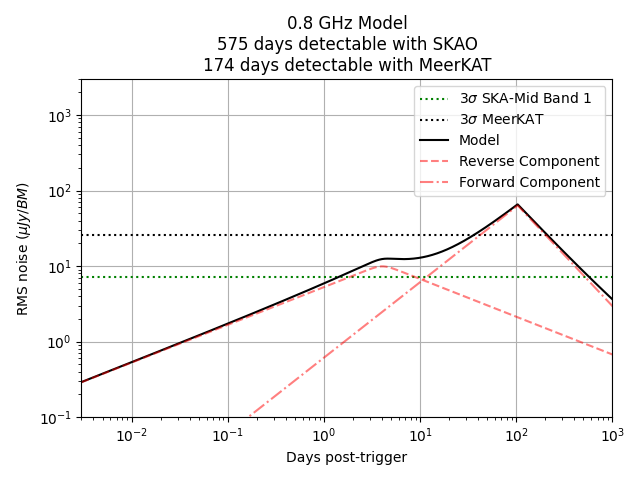}
    \includegraphics[width=0.48\textwidth]{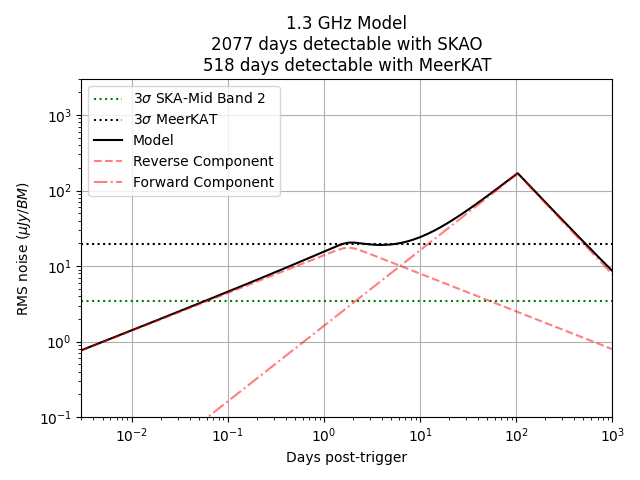}
    \includegraphics[width=0.48\textwidth]{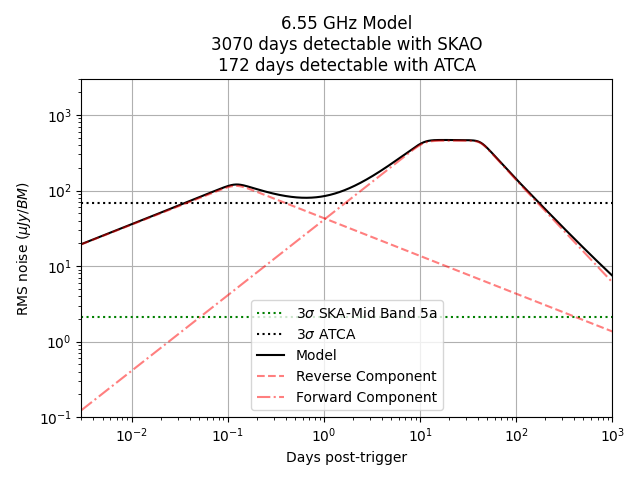}
    \caption{These plots show our models at three frequencies that will be observable with the SKA and compares the sensitivity of current instruments to the Future SKA. }
	\label{fig:lowfreq}
\end{figure}

Our observations between 1.3 and 1.6 GHz were very valuable in constraining the external medium of the burst. This can most clearly be seen in the top left pane of Figure \ref{fig:lcISM}, which shows a large discrepancy between the model and the data, leading us to find that a model with stellar wind type of environment, shown in Figure~\ref{fig:lc}, is the correct density profile. 
We now compare our model and the $3\sigma$ sensitivities for a one hour observation with MeerKAT and ATCA to the expected sensitivity of a one hour integration with SKAO in the corresponding observing band. 
The top panel of Figure~\ref{fig:lowfreq} shows our model along with the current 3~$\sigma$ limit from our one-hour single MeerKAT UHF observation at 0.8\,GHz
and the expected 3-$\sigma$ limit SKA-Mid in Band 1. The middle panel shows the same comparison between our MeerKAT L band (1.3\,GHz) limits and SKA-Mid Band 2. The bottom panel of Figure~\ref{fig:lowfreq} shows our model along with the current 3-$\sigma$ limits from ATCA at 5.5 GHz and the expected 3-$\sigma$ limits with SKA-Mid in Band 5a. The model at 0.8 GHz suggests that we may have been close to detecting the burst and should have continued monitoring the source. Comparing the model light curve with the 1.3 GHz 3-$\sigma$ limit from MeerKAT, if we had continued to take observations we might have detected the afterglow out to a year post-burst or longer. Note that our model does not account for a jet break or for any transition into the non-relativistic phase. Taking observations out to such late times would likely uncover these features and further constrain the dynamics and geometry of the jet. Using current facilities, such as MeerKAT, we may be able to detect bursts out to years and detect changes in the dynamics of the jet. With the future SKAO, we expect to be able to robustly detect these changes at multiple frequencies, potentially allowing for a detailed modeling of the non-relativistic phase.

\subsection{Constraining Physical Parameters}

In our modeling, we are able to detect the reverse shock, constrain $\nu_{a,0,fwd}$, $\nu_{m,0,fwd}$ and $F_{0,fwd}$ for the forward shock, as well as the stellar wind structure of the medium in which the jet is propagating (see Section~\ref{sec:modeling}). However, because we are not able to constrain $\nu_c$, this is not sufficient to fully determine the macrophysics and microphysics of the jet and its environment, e.g., the energy of the blast wave, the magnetic field strength, and the fraction of the shock energy that goes into the radiating electrons. Regarding the latter, we try to obtain some further constraints by using the peaks in the radio light curves and/or spectra, following the methodology described in \cite{2017MNRAS.472.3161B}, \cite{2023MNRAS.518.1522D} and \cite{chastain24}. Since the peak of the afterglow light curve is expected to be $\nu_m$ and there exists certain physical limits expected for the microphysical parameters, they are able to use observables such as the source redshift and distance, isotropic equivalent gamma-ray energy, and the peak's flux density, frequency and time, to put constraints on the fraction of shock energy in the electrons, $\epsilon_e$, and the fraction of electrons being accelerated to relativistic velocities, $\xi_e$. 

We use the equations from \cite{2023MNRAS.518.1522D} to calculate the $\Psi$ parameter, which is a proxy for the aforementioned physical parameters related to electron acceleration, i.e., $\Psi\propto\epsilon_e\xi_e^{-3/2}$, where $\epsilon_e$ is the fraction of shock energy in the electrons and $\xi_e$ is the fraction of accelerated electrons. 
We calculate $\Psi$ for both a homogeneous and a wind environment using the peak of our radio light curves at 1.3 to 1.6 GHz, 2.6 to 3.1 GHz, 5.5 GHz, and 9.0 GHz. For the redshift we use $z=0.824$, as reported by \cite{2024GCN.35698....1F} and employ the Ned Wright's Cosmology Calculator \citep{2006PASP..118.1711W} to compute a corresponding luminosity distance of $1.6\times10^{28}$ cm, using $H_0=69.6$, $\Omega_M=0.286$, and $\Omega_{vac}=0.714$ \citep{2014ApJ...794..135B}. For the gamma-ray energy, we use $E_{\gamma,iso}=1.6\times10^{53}$ ergs based on the fluence reported by \cite{2024GCN.35693....1F}. Combining these values with the observed peak flux densities and corresponding times as shown in Table~\ref{tab:obs}, we estimate ranges for $\Psi$, which we show in Table~\ref{tab:psi}.

\begin{table}[h]
		\caption{$\Psi$ values for GRB 240205B, for a stellar wind and a homogeneous (ISM) medium, based on light curve peaks in different radio observing bands. $T_{mid}$ is the midpoint of the date of the observation at which the flux of the light curve is at its maximum ($F_{max}$) for the observing frequency.}
		\centering
	\label{tab:psi}
	\begin{tabular}{lllll}
		$T_{mid}$     & Freq. & $F_{max}$  & Log10($\Psi_{wind}$) & Log10($\Psi_{ISM}$)  \\
        (Days)        & (GHz) & ($\mu$Jy) &               &                             \\
       \hline
       75.1 & 1.3 to 1.6 & $320\pm85$ & -0.65 to -0.53 & -0.14 to -0.05 \\
       75.1 & 2.6 to 3.1 & $413\pm53$ & -0.57 to -0.51 & -0.10 to -0.07 \\
       17.5 & 5.5        & $532\pm53$ & -0.77 to -0.73 & -0.41 to -0.37 \\
       11.2  & 9.0        & $550\pm18$ & -0.76 to -0.74 & -0.49 to -0.48\\
       \hline
	\end{tabular}
\end{table}


Comparing the $\Psi$ values in Table~\ref{tab:psi} with the distributions of $\Psi$ in \cite{2023MNRAS.518.1522D} reveals that at all frequencies, the ranges of calculated $\Psi$ values for GRB 240205B lie within the range of typical $\Psi$ values for GRB afterglows. For the wind medium, which is the preferred model based on our modeling, the $\Psi$ values are closer to the mean of the larger distribution of GRBs given in \cite{2023MNRAS.518.1522D} than for the homogeneous medium, but none of them are outliers. This suggests that $\epsilon_e$ and $\xi_e$ should likely fall within the typical range for GRBs in a stellar wind environment. Attempting to place constraints on the range of values of $\epsilon_e$ and $\xi_e$ reveal that our constraints on these parameters are less constraining than the ranges given in \cite{2023MNRAS.518.1522D}, which were based on a larger sample of afterglows. Any further constraints on $\epsilon_e$ and $\xi_e$ would require additional observations, perhaps at other wavelengths. The slight discrepancies between the $1-3$ GHz observations and the $4.5-10$ GHz observations could be due to interstellar scintillation, but there is not enough information to be certain. Future studies of GRBs with less scintillation may provide tighter constraints on $\Psi$ and get a better handle on potential evolution of the microphysical parameters.



\subsection{Minimum Bulk Lorentz Factor}

The earliest radio detection of any GRB thus far allows us to put constraints on the minimum bulk Lorentz factor $\Gamma_{\rm{min}}$ of the relativistic GRB blast wave within the first tens of minutes after the gamma-ray trigger. This can be done using brightness temperature arguments \citep{1998Natur.395..663K,1999Natur.398..394G}, as was done for other GRBs with early radio detections \citep[e.g.,][]{2014MNRAS.440.2059A,2018MNRAS.473.1512A,2024ApJ...975L..13A,2025arXiv250814650A}. The underlying principle is that synchrotron-emitting radio sources should have brightness temperatures below the inverse-Compton limit, i.e., the energy of the electrons should not be so high that they will lose their energy to inverse-Compton scattering before they can emit synchrotron radiation. This limit is $\sim10^{12}$~K for non-relativistic sources, but increases for relativistic sources as $\Gamma^3$, where $\Gamma$ is the bulk Lorentz factor of the emitting region. This means that for synchrotron-emitting relativistic jet sources with an observed brightness temperature above $\sim10^{12}$~K, such as GRBs and Active Galactic Nuclei, one can estimate a minimum bulk Lorentz factor $\Gamma_{\rm{min}}$ to explain the apparent discrepancy.

\begin{figure}
	\centering
        \includegraphics[trim=0 0 40 0, width=0.49\textwidth]{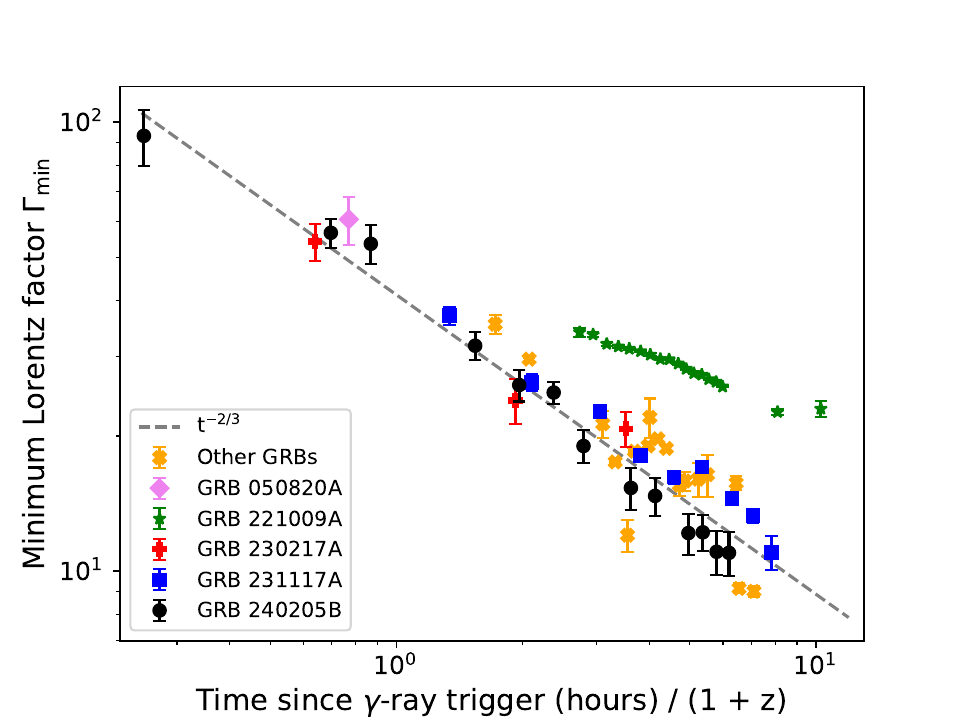}
	\caption{Minimum bulk Lorentz factor $\Gamma_{\rm{min}}$ as a function of 
    cosmological rest frame time for GRBs with detections within the first 12 hours (in observer time) after the gamma-ray trigger \citep{2003AJ....125.2299F,2006ApJ...652..490C,2014MNRAS.440.2059A,2023MNRAS.523.4992A,2024ApJ...975L..13A,2025arXiv250814650A,2015ApJ...815..102F,2021ApJ...906..127F,2015ApJ...810...31V,laskar16grb160509a,2018ApJ...859..134L,laskar19grb190114c,2023NatAs...7..986B}. For times at which there are simultaneous observations in multiple observing bands, only the most-constraining (highest) value for $\Gamma_{\rm{min}}$ is shown. The dashed line indicates the time dependence of the $\Gamma_{\rm{min}}$ calculation to guide the eye and is not a fit.}
	\label{fig:gammamin}
\end{figure}

These arguments have been applied to GRBs with radio detections within the first half a day to one day after the GRB trigger, and the ATCA automated triggering capability has pushed these lower limits on $\Gamma_{\rm{min}}$ to several tens \citep[][]{2024ApJ...975L..13A,2025arXiv250814650A}. Following the same methodology as laid out in the aforementioned publications, we calculate $\Gamma_{\rm{min}}$ using:
\begin{equation}
\Gamma_{\rm{min}}=446.4\cdot d_{\rm{L},28}^{2/3}\cdot F_{\nu,\rm{mJy}}^{1/3}\cdot \nu_{\rm{GHz}}^{-2/3}\cdot (1+z)^{-1/3}\cdot t_{\rm{hr}}^{-2/3}
\end{equation}
with $d_{\rm{L},28}$ the luminosity density in $10^{28}$~cm, $F_{\nu,\rm{mJy}}$ the observed flux in mJy, $\nu_{\rm{GHz}}$ the observing frequency in GHz, $z$ the redshift, and $t_{\rm{hr}}$ the time in hours. The earliest radio detection of GRB 240205B results in a $\Gamma_{\rm{min}}$ value of $\sim10^2$ within the first half an hour after the gamma-ray trigger. This is by far the highest minimum bulk Lorentz factor derived from GRB radio observations. We have compiled all GRBs with radio detections in the first 12 hours, and show the estimated $\Gamma_{\rm{min}}$ as a function of time in Figure~\ref{fig:gammamin}, where the time is redshift-corrected. The figure highlights the three GRBs with early detections in the ATCA automated triggering program, which seem to follow a trend; and almost all GRBs seem to align along this trend, except for GRB 221009A, which was very radio bright \citep{2023NatAs...7..986B}. The trend for most GRBs can be explained by the underlying calculation of $\Gamma_{\rm{min}}$, since Equation~1 shows that $\Gamma_{\rm{min}}\propto t^{-2/3}$. We emphasize that this is the minimum bulk Lorentz factor, so the fact that there is this trend for $\Gamma_{\rm{min}}$ does not mean that the same trend is true for the actual bulk Lorentz factor of the blast wave.

Figure~\ref{fig:gammamin} illustrates the way in which early radio detections can provide insights into GRB jet physics in a model-independent way. Afterglow modeling will give constraints on the blast wave's Lorentz factor as a function of time, but this is always model-dependent. While $\Gamma_{\rm{min}}$ is a lower limit on the Lorentz factor, its estimate using the methodology laid out above is more robust and provides direct input into jet formation and acceleration models. If the trend in $\Gamma_{\rm{min}}$ as a function of time as seen in Figure~\ref{fig:gammamin} holds down to much earlier times, $\Gamma_{\rm{min}}$ would be $\sim1,000$ at 1 minute and even $\sim10^4$ at 1 second after the burst. This can be compared with current $\Gamma_{\rm{min}}$ limits obtained through other methodologies, such as the arrival times of high-energy gamma rays \citep{2009Sci...323.1688A} and gamma-ray variability timescales \citep{2009ApJ...706L.138A}. These studies find $\Gamma_{\rm{min}}$ estimates of $\sim600-1,000$ in the first seconds to minute after the gamma-ray trigger, in line with the extrapolation of early radio detections. More early radio detections within the first hours, or perhaps even minutes, will show what the highest $\Gamma_{\rm{min}}$ value is that GRB jet acceleration models will have to accommodate.

\section{Conclusions}
\label{sec:conclusions}
We have presented the measurements and modeling of an extensive multi-frequency radio observing campaign of GRB 240205B using the ATCA and MeerKAT radio observatories. This data set includes the earliest radio detection of any GRB between 13 and 56 minutes after the gamma-ray trigger. This detection was made by \textit{uv}-fitting 45-minute slices of the initial 12 hour observation. Based on our multi-frequency radio observations, the best-fitting model consists of a forward and reverse shock in a stellar wind environment. This model describes the light curve features over more than an order of magnitude in observing frequency and almost four orders of magnitude in time, with the excursions from the model being consistent with interstellar scintillation. Modeling the data allows us to find the time evolution of the spectral breaks due to synchrotron self-absorption and the minimum electron energy in the forward shock. For the reverse shock, we were able to place limits on the spectral break due to synchrotron self-absorption. 

We have discussed our detection of the reverse shock and promising prospects for future reverse shock detections using rapid-response radio observations. These early radio observations also allow us to constrain the minimum bulk Lorentz factor of the GRB blast wave, and for the first time we estimate this minimum bulk Lorentz factor to be $\sim100$ within the first hour after the GRB trigger. Furthermore, we have discussed the utility and prospects of low-frequency observations (below 1~GHz) of GRB afterglows, and how they can further constrain afterglow models and the structure of the ambient medium.

Based upon our findings, future studies of GRBs using ATCA rapid-response observations promise to detect and better constrain any possible reverse shock emission during the very early afterglow. 
Indeed, conducting rapid-response observations with SKA-Mid or the Karl G. Jansky Very Large Array would provide additional sensitivity on minute timescales that could better constrain the evolution of this synchrotron component. 
Additionally, our model suggests that future late-time observations at low frequencies ($\le 1$ GHz) could provide additional constraints on the profile of the ambient medium, better constraints on $\nu_a$, as well as potentially detect transitions in the jet dynamics from a jet break or from a transition to the non-relativistic regime. Future observations of GRB afterglows will place additional constraints on $\Gamma_{\rm{min}}$ in a model-independent way, and allow for comparison with other methods for constraining the bulk Lorentz factor of the blast wave. For bursts that also have measurements of the rise time of the afterglow at optical wavelengths and/or high-energy gamma-ray observations, it will be possible to place constraints on the early evolution of the GRB jet, including potentially its acceleration phase. 

\section{Acknowledgments}

We acknowledge the Gomeroi people as the traditional owners of the ATCA observatory site. The ATCA is part of the Australia Telescope National Facility (\href{https://ror.org/05qajvd42}{https:// ror.org/ 05qajvd42}), which is funded by the Australian Government for operation as a National Facility managed by CSIRO. 
This paper includes archived data obtained through the Australia Telescope Online Archive ( \href{http://atoa.atnf.csiro.au}{http://atoa.atnf.csiro.au}).

This work made use of data supplied by the UK \textit{Swift} Science Data Centre at the University of Leicester and the Swift satellite.
\textit{Swift}, launched in 2004 November, is a NASA mission in partnership with the Italian Space Agency and the UK Space Agency. \textit{Swift} is managed by NASA Goddard. Penn State University controls science and flight operations from the Mission Operations Center in University Park, Pennsylvania. Los Alamos National Laboratory provides gamma-ray imaging analysis.

The MeerKAT telescope is operated by the South African Radio Astronomy Observatory (SARAO), which is a facility of the National Research Foundation, an agency of the Department of Science and Innovation.

This work made use of the CARTA (Cube Analysis and Rendering Tool for Astronomy) software (DOI 10.5281/zenodo.3377984 –  https://cartavis.github.io)

We would like to thank the UNM Center for Advanced Research Computing, supported in part by the National Science Foundation, for providing the high performance computing resources used in this work.

LR acknowledges support from the Trottier Space Institute Fellowship and from the Canada Excellence Research Chair in Transient Astrophysics (CERC-2022-00009). 

JKL acknowledges support from the University of Toronto and Hebrew University of Jerusalem through the University of Toronto - Hebrew University of Jerusalem Research and Training Alliance program. 
The Dunlap Institute is funded through an endowment established by the David Dunlap family and the University of Toronto. 

\bibliography{sample701}{}
\bibliographystyle{aasjournalv7}



\end{document}